\documentclass[prl,twocolumn,superscriptaddress,showpacs,floatfix]{revtex4}
\usepackage{graphicx}
\usepackage{bm} 
\usepackage{amsmath}
\usepackage{amssymb}
\begin{document}
\title{Dynamics of fluctuations in an optical analogue of the Laval nozzle}

\author{Itzhak Fouxon}

\affiliation{Raymond and Beverly Sackler School of Physics and
Astronomy, Tel-Aviv University, Tel-Aviv 69978, Israel}

\author{Oleg Farberovich}

\affiliation{Raymond and Beverly Sackler School of Physics and
Astronomy, Tel-Aviv University, Tel-Aviv 69978, Israel}

\author{Shimshon Bar-Ad}

\affiliation{Raymond and Beverly Sackler School of Physics and
Astronomy, Tel-Aviv University, Tel-Aviv 69978, Israel}

\author{Victor Fleurov}

\affiliation{Raymond and Beverly Sackler School of Physics and
Astronomy, Tel-Aviv University, Tel-Aviv 69978, Israel}


\date{\today}

\begin{abstract}
Using the analogy between the description of coherent light
propagation in a medium with Kerr nonlinearity by means of nonlinear
Schr\"odinger equation and that of a dissipationless liquid, we
propose an optical analogue of the Laval nozzle. The optical Laval
nozzle will allow one to form a transonic flow, in which one can
observe and study very unusual dynamics of classical and quantum
fluctuations, including an analogue of the Hawking radiation of real
black holes. Theoretical analysis of these dynamics is supported by
numerical calculations, and estimates for a possible experimental
realization are presented.
\end{abstract}


\maketitle

Black hole radiation is one of the most impressive phenomena at the
intersection of general relativity and quantum field theory.
Accounting for the quantum nature of the physical vacuum led to the
prediction that a black hole, defined classically as an object that
even light cannot escape, in fact can be characterized by a
temperature, entropy\cite{Bekenstein} and moreover emits thermal
radiation\cite{H75}. Since direct experimentation with black holes
is hardly possible, it was suggested to consider analogous phenomena
in condensed matter physics where the "high-energy"
(short-wavelength) physics is known \cite{U81}. The suggestion was
based on the observation that the derivation of the Hawking
radiation uses only the linear wave equation in curved space-time,
and not the Einstein equations. The same conditions for wave
propagation arise when considering sound propagation in a fluid when
the background flow is non-trivial \cite{U81}. This similarity was
shown to be sufficient for applying the original considerations by
Hawking, predicting the existence of thermal radiation of quantum
origin from the fluid counterpart of the horizon, which can be
called Mach horizon (where the fluid velocity equals the sound
velocity, \emph{i.e.} Mach number $M=1$). These ideas were further
developed for possible applications in BEC fluids and other systems
\cite{R00,BLV03,CFRBF08,RPC09,NBRB09}. As for experimental
realizations, a white hole horizon was observed in optical
fibers\cite{PKRHKL08}, where the probe light was back-reflected from
a moving soliton, and a black hole horizon was observed in a BEC
system\cite{LIBGS09}.

We propose an experimental setup capable of creating a Mach horizon
in an optical medium with Kerr nonlinearity, which may be called
optical Laval nozzle. The propagation of weakly nonlinear coherent
optical pulses is described by the non-linear Schr\"odinger (NLS)
equation
\begin{equation}\label{NLS}
i\frac{\partial A}{\partial z} = -\frac{1}{2 \beta_0}\nabla^2A +
U(x,y)A + \lambda |A|^2 A.
\end{equation}
It assumes the paraxial approximation, according to which the
electric field of the light wave is written as $E = A(x,y,t;z)
e^{-i\beta_0 z}$, where $A(x,y,t;z)$ is weakly $z$ dependent complex
amplitude of the light propagating in the $z$ direction. The time
coordinate $t$ is converted into the $\tau$ coordinate\cite{A95}
which describes the shape of the light pulse in the moving
coordinate system. The radius vector differential is now $d{\bm r} =
(dx,dy,d\tau)$ (anomalous dispersion) or $d{\bm r} = (dx,dy,-i
d\tau)$ (normal dispersion). $U(x,y)$ is the equivalent external
potential created by spacial variation of the refraction index in
the medium and does not depend on $\tau$.

Now we briefly sketch the fundamental result\cite{U81} as applied to
coherent light propagation in a medium with Kerr nonlinearity. The
Madelung transformation\cite{M27} $A = f e^{i\phi}$ (see
also\cite{M75,DFSS07,Marino08}) maps the NLS eq.(\ref{NLS}) on two
equations
\begin{eqnarray}
&&
\partial_z\rho + \nabla\cdot[\rho{\bm v}] = 0,\label{b1}
\\
&& \partial_z\bm v + \frac{1}{2} \nabla{\bm v}^2 = -
\frac{1}{\beta_0}\nabla\left( V_{qu} + U + \lambda \rho
\right)\label{b2}
\end{eqnarray}
for an equivalent fluid with density $\rho = f^2$ and velocity
$\beta_0{\bm v} = -\nabla\phi$. For permanently shined light the
coordinate $\tau$ is redundant. $V_{qu} = - \frac{1}{2\beta_0}
\frac{\nabla^2 f}{f}$ is a "quantum potential" corresponding to
$\hbar =1$. The light wave vector $\beta_0$ is the "particle mass"
and the velocity $v$ is dimensionless.

Linearizing Eqs. (\ref{b1}) and (\ref{b2}) with respect to the small
fluctuations $\rho - \rho_0 = \rho_0\psi$ and $\phi - \phi_0 =
\varphi$ around a steady solution $\rho_0(x,y)$ and $\varphi_0(x,y)$,
we arrive at
\begin{equation}
(-g)^{-1/2}
\partial_{\mu}(-g)^{1/2} g^{\mu\nu} \partial_{\nu} \varphi = 0,
\label{b4}
\end{equation}
where the metric $g^{\mu\nu}$ with the determinant $g$ is given by
the interval
\begin{equation}\label{b5}
d\sigma^2 = g_{\mu\nu} dx^\mu dx^\nu =
$$$$
\sqrt{\frac{\beta_0}{\lambda\rho_0}}\left[d{\bm r}^2 - 2dz{\bm
v}_0\cdot d{\bm r}- \left(\frac{\lambda\rho_0}{\beta_0} -
v_0^2\right) dz^2\right]
\end{equation}
with $g^{\mu\mu'}g_{\mu'\nu} = \delta^\mu_\nu$. The ordinary "sound
waves" in the effective fluid arise as solutions of eq. (\ref{b4})
around the equilibrium solution $\rho_0 = const$, $\phi_0=const$ and
$\bm v_0 =0$, that exists at $U=0$, so that $\lambda\rho_0/\beta_0 =
s^2$ is the squared sound velocity. The nonlinearity coefficient is
positive, $ \lambda > 0$, otherwise the "sound velocity" becomes
imaginary and various instabilities, such as collapsing
solitons\cite{S90} arise. Eq. (\ref{b4}) may include corrections due
to the quantum potential, which may be of importance for short
waves.

The background flow may arise in the Laval nozzle\cite{S65,LL87},
\emph{i.e.} a vessel with variable cross-section $S(x)$, whose
application to some condensed-matter analogues was discussed
in\cite{SakagamiOhashi,BLV03a}. The flow velocity increases with
decreasing $S(x)$, and reaches the sound velocity in the narrowest
part of the vessel, called the throat. Further acceleration is
reached by increasing $S(x)$. Here we analyze the transonic flow for
the optical realization of the Laval nozzle shown in Fig.
\ref{Laval}.
\begin{figure}[ht]
 \includegraphics[width=8cm,angle=-0]{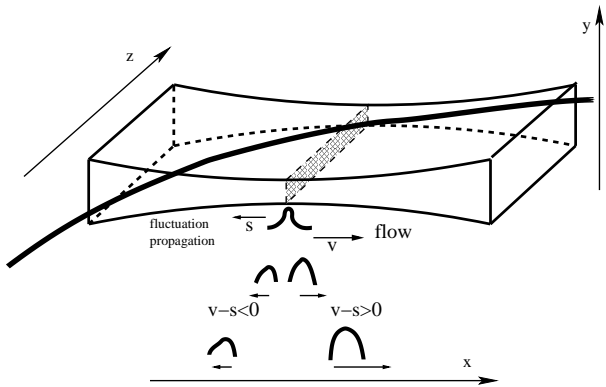}
\caption{A cartoon of a laser beam propagating inside the optical
Laval nozzle. The beam, with a small initial angle relative to the z
axis, bends more towards the x direction as it propagates along the
z direction ("time"). The bend in the cartoon is strongly
exaggerated. A fluctuation is schematically shown near the Mach
horizon as it is cut in two parts which propagate in opposite
directions. } 
\end{figure}
The laser beam, nearly parallel to the $z$ axis, is tilted in the
$x$ direction by an angle $\arctan (k_x/\beta_0)$, \emph{i.e.} the
initial phase in the electric field amplitude $A(x,y)$ is $\varphi =
k_x x$. $k_x/\beta_0$ plays the part of "flow velocity" in the $x$
direction. Acceleration of the flow within the plate of variable
width ($y$ - axis) results in bending of the beam ($k_x$ increases)
as shown in  Fig. \ref{Laval}. It leads to small corrections to
$\beta_0$, quadratic in $k_x/\beta_0 \ll 1$. We always assume that
the wave length $1/\beta_0$ is smaller than all the other relevant
scales.

Now we consider the flow bounded by the walls of a hyperbolic shape
\begin{equation}\label{hyperbola}
\frac{y^2}{a^2} - \frac{x^2}{b^2} = 1
\end{equation}
with sliding boundary conditions, and the elliptic coordinates
$
x = c \sinh u_1 \sin u_2, \ \ y = - c \cosh u_1 \cos u_2,
$
are introduced, with $c =\sqrt{a^2 + b^2}$. From the symmetry
considerations the flow is along the x axis for $y = 0$, \emph{i.e.}
$u_2 = \pi/2$.

Equation (\ref{b1}) for a steady flow becomes
$$\frac{\partial}{\partial u_1} \sqrt{\Delta} (\rho v_1) +
\frac{\partial}{\partial u_2} \sqrt{\Delta}(\rho v_2)  = 0
$$
where $\Delta = \sinh^2u_1 + \sin^2 u_2$ and $v_1$ and $v_2$ are the
velocity components in the elliptic coordinates.

This equation is satisfied if
\begin{equation}\label{velocity}
\begin{array}{c}
\rho v_1 = \displaystyle \frac{F_1(u_2) + f u_1}{\sqrt{\Delta}},
\\
\rho v_2 = \displaystyle \frac{F_2(u_1) - f u_2}{\sqrt{\Delta} }
\end{array}
\end{equation}
where $F_1(u_2)$ and $F_2(u_1)$ are functions only of $u_2$ and
$u_1$ respectively, and $f$ is a constant. Since the transversal
velocity $v_2$ is zero at the throat at $u_1=0$ we get that $f=0$.

We assume the streamlines in the vicinity of the central line,
\emph{i.e.} small $\widetilde{u}_2 = u_2 -
\displaystyle\frac{\pi}{2}$, to be close to the lines of constant
$\widetilde{u}_2$. Then calculating the covariant derivatives in the
curvilinear elliptic coordinates and neglecting in the result the
second order corrections in $\widetilde{u}_2$, Eq. (\ref{b2}) reads
(cf., \cite{LL87})
\begin{equation}\label{velocity_1}
\frac{\partial (\rho v_1)}{\partial u_1} = \rho \left(1 -
\frac{v_1^2\beta_0}{\lambda\rho}\right) \frac{\partial v_1}{\partial
u_1}.
\end{equation}
The flow reaches the sound velocity at $u_1 = 0$. The second Eq.
(\ref{velocity}) in the elliptic coordinates results in
$ \rho(u_1) v_1(u_1) = d - \frac{1}{2} d u_1^2 $
with $ d = F_1(\pi/2) = \overline{\rho}\ \overline{s} $, and
finally, using Eq. (\ref{velocity_1}), we get
$v_1 = \overline{s} (1 + \frac{u_1}{\sqrt{3}}), \ \ \ \rho =
\overline{\rho} (1 - \frac{u_1}{\sqrt{3}}) $,
where overlined quantities refer to the throat. The coordinate
dependence of the density immediately yields the coordinate
dependence of the sound velocity $\lambda \rho(u_1) = \beta_0
s^2(u_1)$. Correspondingly the values of these quantities at the
throat are connected as $\lambda \overline{\rho} = \beta_0
{\overline{s}}^2$.

We may now represent Eq. (\ref{b4}) in the narrow transonic region
in the dimensionless form
\begin{equation}\label{KG}
\partial^2_{\zeta}\varphi + 2
\partial_\xi\partial_\zeta \varphi + \alpha \xi
\partial^2_\xi\varphi + \alpha_1
\partial_\zeta \varphi + \alpha \partial_\xi \varphi
=0, 
\end{equation}
where $ \zeta = \displaystyle \frac{ \overline{s}}{c} z,\ \ \ \xi =
u_1 \approx \displaystyle \frac{x}{c}$. The coefficients $\alpha_1 =
1/\sqrt{3}$, and $\alpha = \sqrt{3}$ are obtained above for the
hyperbolic Laval nozzle. However, the ratio $\alpha/\alpha_1 = 3$
holds for any differentiable shape of this type. We look for a
solution of Eq. (\ref{b9}) in the form
$\varphi(\xi,\zeta) = \exp\{i\nu\zeta\} \overline{\varphi}(\xi) $,
where the functions $\overline{\varphi}(\xi)$ are linear
combinations of the two sets of eigen functions
\begin{equation}\label{eigen-1}
\overline{\varphi}_1(\xi) = e^{\frac{-i\nu}{\alpha}\mbox{ln}(-\xi)}
%
\ \ \mbox{and}\ \ \overline{\varphi}_2(\xi) = e^{-ik(\nu)\xi}.
\end{equation}
In a background flow characterized by a space independent velocity,
eq. (\ref{KG}) has the form of a one-dimensional Klein-Gordon wave
equation for the field $\varphi$ in a flat space. Then one obtains
two plane waves propagating to the left (against the flow) and to
the right (with the flow). In the considered here transonic case,
when the background flow velocity and speed of sound vary with the
coordinate, the left moving plane wave transforms into the eigen
mode $\overline{\varphi}_1(\xi)$, which is a function with a
branching point near $\xi = 0$, whereas the right moving plane wave
is characterized by the nonlinear spectral relation
\begin{equation}\label{spectrum}
k(\nu) = \frac{\nu}{2} \frac{1 - i/(\nu\sqrt{3})}{1 -
i\sqrt{3}/(2\nu) }.
\end{equation}
The complex wave vector for the real "frequencies" reflects the fact
that the coordinate $\xi$ of these eigen modes corresponds to motion
along diverging hyperbolic streamlines. However, the frequency $\nu$
is real which indicates linear stability of the flow with respect to
weak fluctuations.

The corresponding eigen modes
\begin{equation}\label{eigen-2}
\overline{\psi}_1(\xi) = - i\nu
\frac{\overline{s}}{c\overline{\rho}} \frac{1 + \alpha\xi
}{\alpha\xi} \overline{\varphi}_1(\xi),
%
$$$$
\overline{\psi}_2(\xi) = - i(\nu - k(\nu))
\frac{\overline{s}}{c\overline{\rho}} \overline{\varphi}_2(\xi)
\end{equation}
for the density fluctuations can be obtained directly from the
linearization of the Euler equation (\ref{b2}).

Crossing the Mach horizon at $\xi=0$, the solution
$\overline{\varphi}_1(\xi)$ acquires the factor $
\exp\{\frac{\nu\pi}{\sqrt{3}} \}$. The same factor appears when
considering quantum fluctuations. Following the logic of \cite{H75}
(also discussed in detail in \cite{DL08} --- eqs. (2.88)-(2.90))
this results in radiation to the left of the Mach horizon (outside
the "black hole") with the frequency distribution $N(\nu) =
1/(\exp\{\frac{\hbar \nu}{T_H} \} - 1)$. This distribution has the
same form as a Planck distribution for black body radiation, with
the parameter
$ T_H = \frac{\sqrt{3} \overline{s} \hbar}{2\pi c } $
in physical dimensions. The latter plays a role similar to that of
the Hawking temperature in black hole radiation. However the
quantity $T_H$ is not a real physical temperature. The part of time
is played here by the spatial coordinate $z$, \emph{i.e.} the
"frequency" $\nu$ is the wave vector of a fluctuation in the $z$
direction. This "temperature" is measured in units of momentum
rather than energy, and the corresponding "thermal fluctuations" are
waves along the $z$ direction with a typical wavelength of order
$\lambda_H = \hbar/T_H = 2\pi c/(\sqrt{3} s)$.

A "straddled" fluctuation
\begin{widetext}
\begin{equation}\label{wavepacket}
f(\xi,\zeta) = \int d \nu g(\nu) e^{-i\nu\zeta}
\overline{\varphi}_1(\xi) =
$$$$
\exp\left \{ i\left[\nu_0 + \sqrt{3} \Gamma^2 \pi \theta(\xi)
\right] \left(\sqrt{3} \ln |\xi| + \zeta\right)\right\} \exp\left\{
\left[\frac{\pi\nu_0}{\sqrt{3}} + \frac{\Gamma^2\pi^2}{6} \right]
\theta(\xi)\right\} G(\xi,\zeta)
\end{equation}
\end{widetext}
($\theta(\xi)$ is the step function) with a Gaussian spectral
density
$
g(\nu) = \frac{1}{\Gamma\sqrt{2\pi}} \exp\left\{- \frac{(\nu -
\nu_0)^2}{2\Gamma^2}\right\}
$
around a positive frequency $\nu_0$ of a width $\Gamma$ is composed
exclusively of the "left moving" normal modes
$\overline{\varphi}_1(\xi)$, singular at $\xi = 0$. Its "time"
evolution (\emph{i.e.} $\zeta$ dependence) is determined by the
envelope function
$
G(\xi,\zeta) = \exp\left\{- \frac{\Gamma^2}{2} \left(\sqrt{3} \ln
|\xi| - \zeta\right)^2 \right\}.
$
The "frequencies" of these oscillations on both sides of the horizon
strongly vary with $\xi$. This is reminiscent of "time slowing" near
the horizon of a real black hole.

\begin{figure}[ht]
\includegraphics[width=6cm,angle=-90]{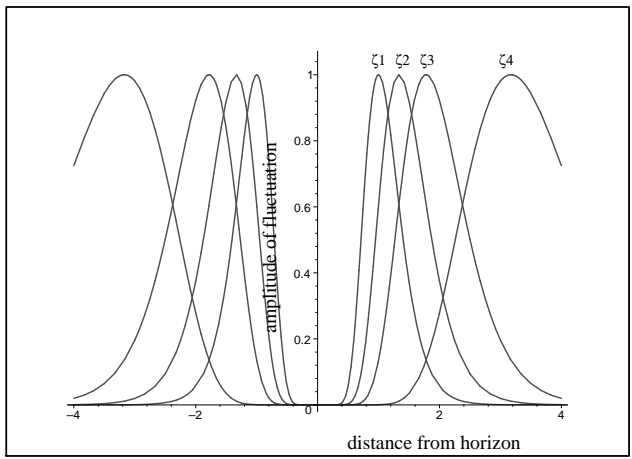}
\caption{Propagation of the straddled wave packet to the left and
right of the Mach horizon. $\zeta_1 = 0$, $\zeta_2 = 0.5$, $\zeta_3=
1$, $\zeta_4 = 2$ with $\Gamma= 2$.}\label{fluctuation-1}
\end{figure}

Fig. \ref{fluctuation-1} shows how the maxima of the fluctuation
amplitude propagate on both sides of the Mach horizon, with zero
amplitude on the horizon. The part residing to the right side of the
horizon (the supersonic region, $\xi > 0 $) should be multiplied by
the factor
$ \exp\left\{\frac{\pi\nu_0}{\sqrt{3}} +
\frac{3\Gamma^2\pi^2}{2}\right\} > 1$
(not shown in Fig. \ref{fluctuation-1}), with the Hawking
temperature $\pi\nu_0/\sqrt{3} = \hbar\nu_0/2T_H$ as a parameter.
Therefore the ratio of amplitudes of a classical fluctuation on the
left and the right of the horizon directly measures the Hawking
temperature.

The fluctuation moves against the transonic flow with the velocity
of sound. Its left part escapes to the left against the subsonic
background flow, and  moves with a very small velocity $s_l - v_l
\approx \overline{s} - v_l$ to the left. The right part of the
fluctuation is "flushed" away by the supersonic flow (see Fig.
\ref{Laval}), and moves to the right with a small velocity $v_r -
s_r \approx v_r - \overline{s}$. Here $v_l$ and $s$ are the subsonic
flow velocity and local speed of sound of the flow in the vicinity
and to the left of the Mach horizon, whereas $v_r$ and $s_r$ are the
corresponding quantities in the supersonic region to the right of
the Mach horizon (Note that for the symmetrical potential and
spectral content of the fluctuation studied here the propagation on
the two sides of the horizon is approximately symmetrical, as shown
in Fig. \ref{fluctuation-1}). Using the eigen modes (\ref{eigen-2})
for the density fluctuations we get a similar picture. In the case
of quantum fluctuations the part moving to the left is analogous to
the Hawking radiation observable outside the black hole. The part
moving to the right is not observable in a real black hole. However
in our case the supersonic region is accessible for observation.

The normal modes $\overline{\varphi}_2(\xi)$ in (\ref{eigen-1}) are
plane waves propagating to the right in the direction of the flow
with the velocity
$
v_2 = \overline{s} \left(\frac{d k(\nu)}{d\nu}\right)^{-1}.
$
This velocity is twice the sound velocity, \emph{i.e.} the sound
velocity on the background of the supersonic flow, for large enough
"frequencies" $\nu \gg 1$. At small frequencies $\nu \ll 1$ the
propagation velocity is closer to thrice the sound velocity,
\emph{i.e.} twice the sound velocity relative to the background
flow. If we use the analogy with the real black holes this result
would correspond to superluminal motion. However there is no paradox
here since the wavelength is larger than the throat width in the
Laval nozzle, $1/k \gg c$ (in physical units). Hence we deal with
the narrow field effect, analogous to superluminal propagation in
near field optics and tunneling \cite{BK01,MRR00,MRS98}, which does
not violate causality.

The above results are illustrated by numerical simulations of the
NLS eq. (\ref{NLS}) in 1 + 2 dimensions using a finite-difference
approach based on the split-step Crank-Nicholson method.\cite{MA09}
The potential $U(x,y)$ is zero outside the hyperbola
(\ref{hyperbola}) and a negative constant $U_0$ inside. A packet
with a Gaussian density distribution is given a Galilean boost in
the fashion similar to Ref.\cite{FPR92} with a velocity close to
that of sound for the density in the maximum of the Gaussian. This
configuration differs from the discussed in the analytical part of
the paper and is chosen for its relative simplicity and due to the
fact its experimental realization seems to be rather
straightforward. Although it differs in many aspect such as strongly
nonhomogeneous distribution of the density and, hence, of the sound
velocity, which is also affected be the quantum potential
contribution, this configuration may be very convenient and useful
for studying propagation and {\em classical} fluctuations in the
optical Laval nozzle.

\begin{figure*}[ht]
 \includegraphics[width=8cm,angle=-0]{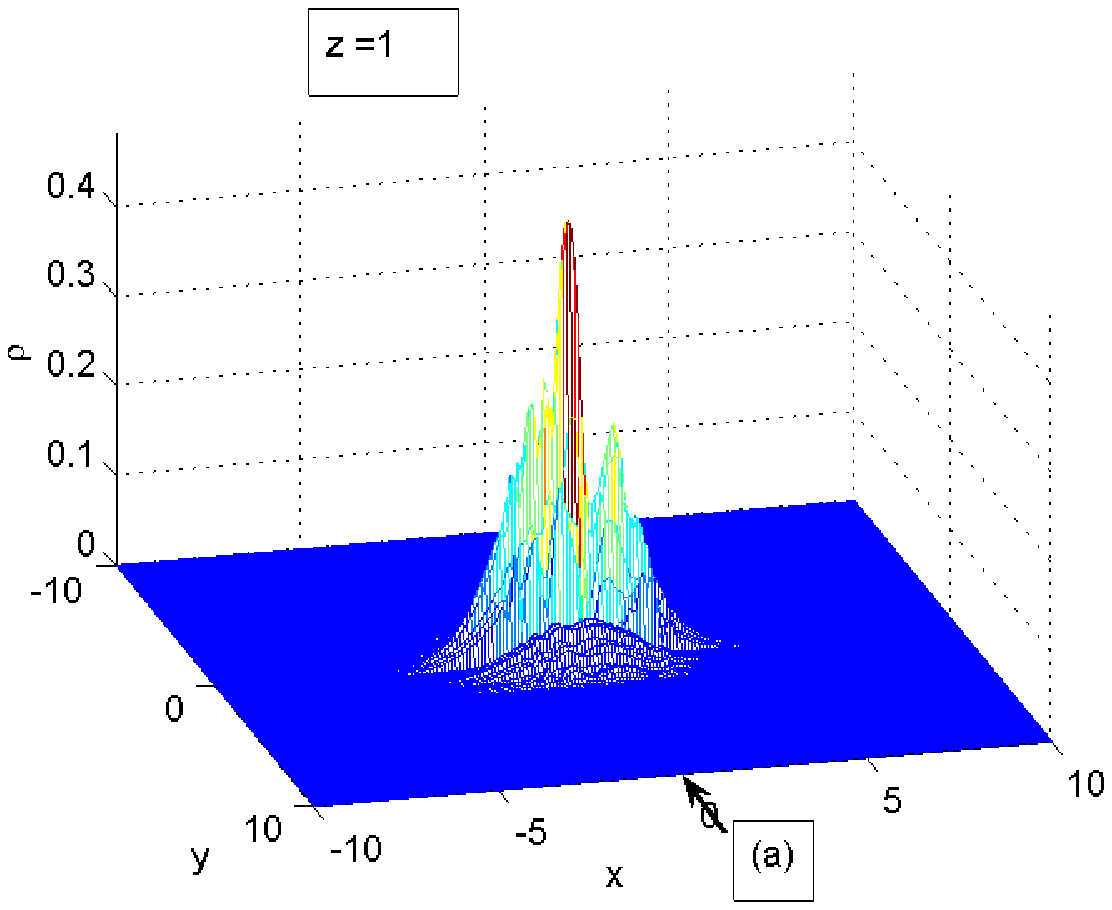}
 \includegraphics[width=8cm,angle=-0]{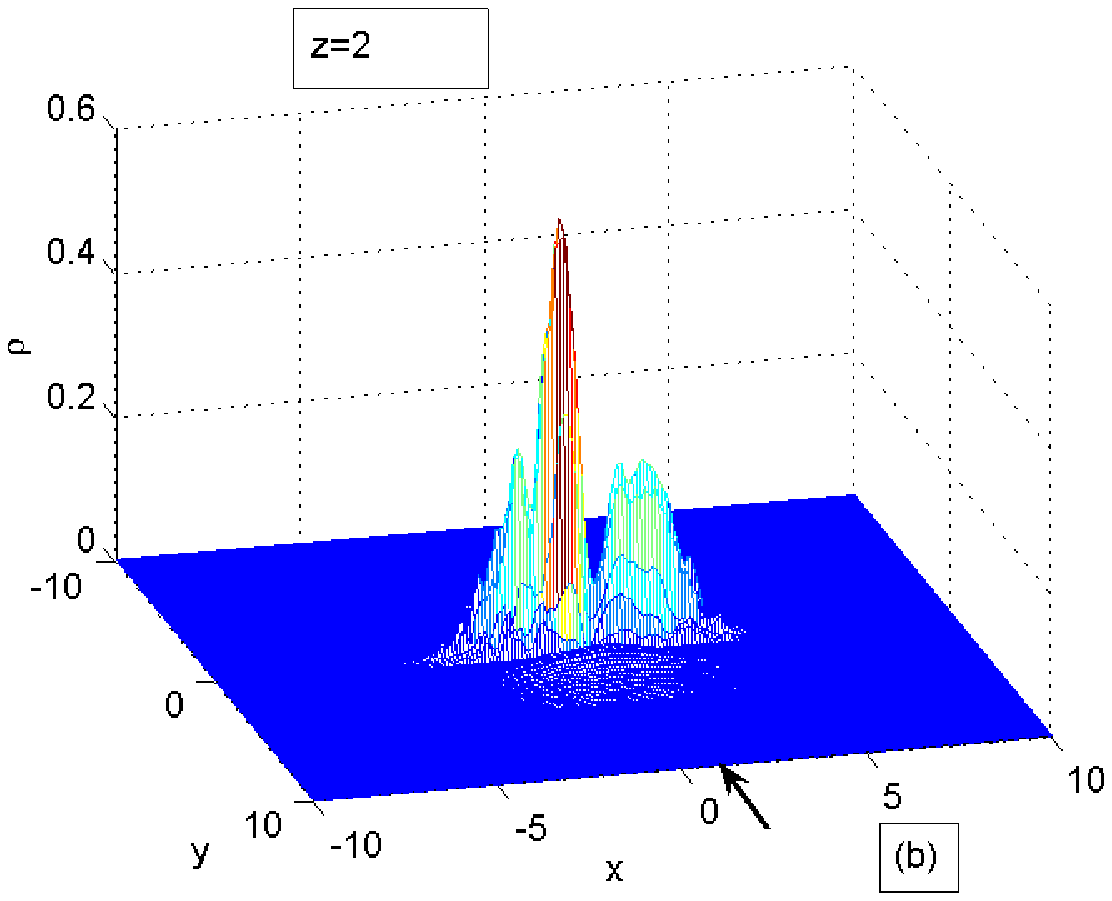}
 \includegraphics[width=8cm,angle=-0]{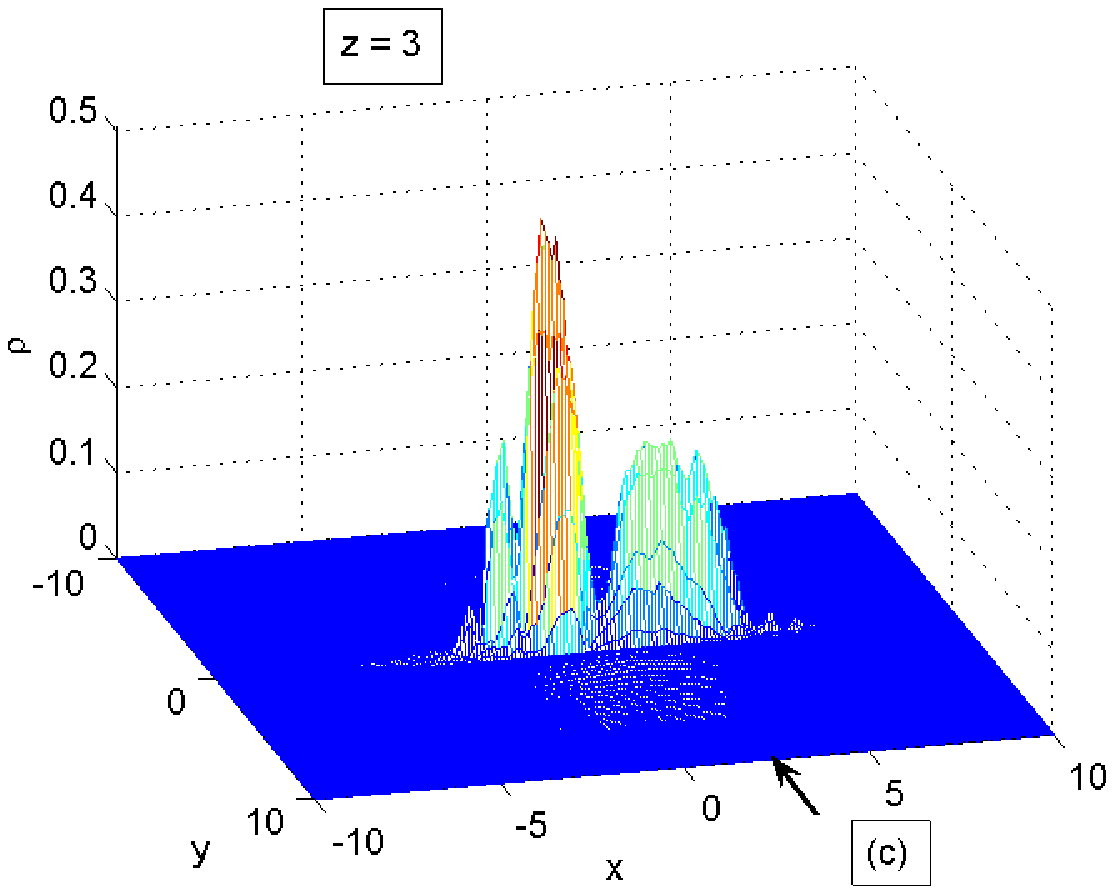}
 \includegraphics[width=8cm,angle=-0]{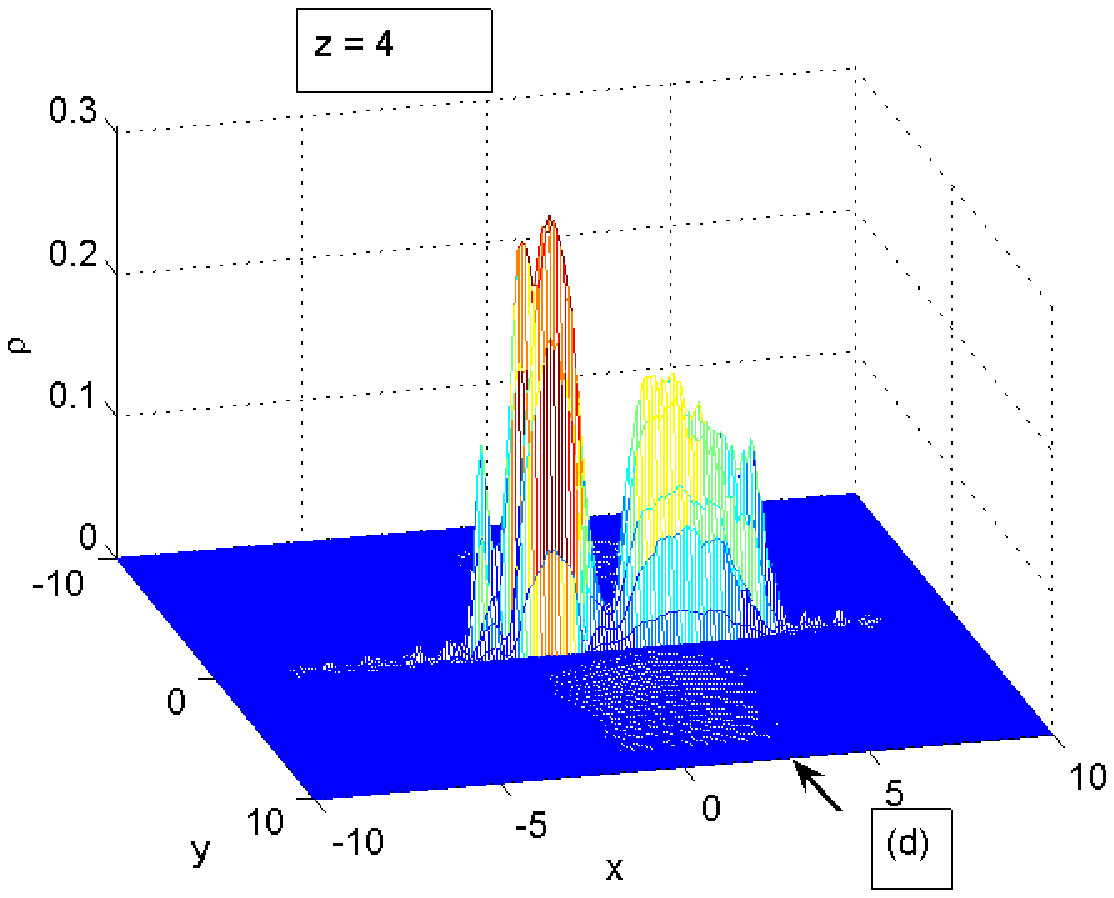}
\caption{(Color online) Four figures (a) to (b) show snapshots at
different stages of the Gaussian packet motion through a hyperbolic
Laval nozzle and development a straddled fluctuation with a
characteristic dip close to the throat of the nozzle. The arrow in
figure (a) to (d) shows the center of the packet if it were moving
in a free space, which corresponds to the shift $\Delta x = 1,\ 2,\
3$ and 4, respectively, from the starting point $x =-1$. The
propagation distance $z$ ("time") is indicated at each
figure.}\label{straddle}
\end{figure*}

The calculation is carried out for Eq. (\ref{NLS}) with $\beta_0 =
5$ and $\lambda = 10$, whereas the amplitude function $A$ is
normalized to unity. The depth of the potential in the nozzle $U_0 =
-200$. It has a hyperbolic shape with $a = 0.3$ and $b = 7.5$.
Meaning that its half-width is 0.3 at the throat at $x = 0$ and 0.5
at $x = \pm 10$. The Gaussian centered initially at $x = -1,\ y=0$
is given a boost with the phase $\varphi = 2.5x $ corresponding to
the unit velocity. Figs. \ref{straddle} show the packet at four
successive stages of its propagation. A straddled fluctuation
gradually develops with a characteristic dip close to the throat of
the nozzle, which becomes deeper and nearly cuts the original packet
in two. The fluctuation in this case becomes strong and in the end
consumes a major part of the packet.

Since the numerical code deals with the NLS eq. (\ref{NLS}), it
solves a nonlinear problem strongly deviating from the steady
transonic flow assumed in the above analysis, and fully accounts for
the quantum potential. The density is strongly nonhomogeneous which
leads to a nonhomogeneous distribution of the sound velocity. The
latter should be also corrected by the contribution of quantum
potential which cannot be considered small anymore. The Gaussian
packet evolves into a two hump structure permanently changing its
shape with the propagation length $z$ ("time"), as is clearly
demonstrated in Figs. \ref{straddle}.

There are such factors as quantum potential (dispersion) and
defocusing nonlinearity, which may influence the dynamics of the
packet and in principle also cause its splitting in two parts.
Control calculations of the packet moving in free space, or in a
potential well with parallel walls do not show the type of behavior
shown in Fig. \ref{straddle}. Decreasing the boost velocity also
weakens the effect. Otherwise the general pattern presented in these
figures is rather robust and holds when varying the parameters as
long as the packet is boosted in a hyperbolically shaped potential
well.F

The above ideas can be tested experimentally by launching a
continuous wave laser beam through an appropriately shaped nozzle,
with reflective walls, filled with a Kerr-like defocusing nonlinear
material. Such a one-dimensional nozzle can be constructed from two
convex cylindrical mirrors, put back to back at an adjustable
separation with the space between them sealed and filled with a
nonlinear liquid such as iodine-doped ethanol\cite{Jason}. The
nonlinear coefficient $\lambda$ of the solution can be expressed, in
terms of the nonlinearly-induced refractive index change $\delta n$,
as $\delta n \beta_0 / n_0$, where $n_0$ is the linear refractive
index of the material\cite{Jason}. The corresponding dimensionless
sound velocity is $s = \sqrt{\delta n / n_0}$. Assuming a nonlinear
index change of 1\%, an incoming laser beam would emulate supersonic
flow when $k_x / \beta_0 > 0.1$, \emph{i.e.} when the input angle of
the beam is larger than $\approx 6$ degrees. The required input
angle for observing transonic acceleration through the nozzle is
therefore technically feasible. The acceleration can be controlled
by mechanically varying the throat width, and monitored by observing
the deflection of the laser beam as it traverses the nozzle. The
intensity and phase profile of the beam coming out of the
experimental setup can be studied using standard detection and
interferometric techniques. These would allow measurement of the
transverse acceleration of the beam, as well as detection and
characterization of classical fluctuations that propagate away from
the throat region. such classical fluctuations may be introduced by
locally perturbing the optical fluid in a deterministic fashion,
\textsl{e.g.} by injecting a second narrow laser beam into the
nozzle throat\cite{Marino08}.

Constructing a Laval nozzle of hyperbolic shape might not be easy,
but we believe that most of the effects discussed above will also be
observed in other types of convergent-divergent nozzles (in fact
many aeronautical applications of the Laval nozzle are not
hyperbolically shaped). Otherwise the proposed experimental setup is
quite simple and straightforward to implement in a table-top
experiment, in particular as compared to the previously proposed
rotating black hole configuration\cite{Marino08} and to water tank
experiments\cite{Rousseaux08}. The (classical) fluctuations will be
observed directly, superimposed on an otherwise smooth transverse
beam profile, and in principle will not require as high a dynamic
range as previous optical experiments\cite{PKRHKL08}. The experiment
is also more straightforward than other proposed schemes,
\textsl{e.g.} those based on SQUID array transmission lines
\cite{NBRB09} and atomic BEC\cite{CFRBF08,RPC09}.

In summary, we demonstrate the possibility of creating an optical
analogue of the Laval nozzle. The interesting point will be to study
the dynamics of straddled fluctuations which may be either quantum
or classical and even artificially created. The equivalent of the
Hawking temperature enters as an important parameter characterizing
all types of fluctuations. This temperature is measured in units of
momentum, rather than energy, which corresponds to a wavelength
exceeding the width of nozzle throat by about two orders of
magnitude, and in principle accessible for experimental
measurements. Numerical simulations support the theoretical findings
but also indicate that the phenomenon is much more complicated and
interesting.

{{\bf Acknowledgments.}\ \  Support of Israeli Science Foundation,
Grant N 944/05 and of United States - Israel Binational Science
Foundation, Grant N 2006242 is acknowledged. VF is grateful to Max
Planck Institute for Physics of Complex Systems, Dresden, for
hospitality and to G. Shlyapnikov and N. Pavloff for stimulating
discussions.}

\end{document}